\documentclass[aps,showpacs]{revtex4}
\usepackage{graphicx}
\newcommand{\be}{\begin{equation}}
\newcommand{\ee}{\end{equation}}
\newcommand{\ben}{\begin{eqnarray}}
\newcommand{\een}{\end{eqnarray}}
\newcommand{\ov}{\overline}
\begin{document}

\title{Kinks and domain walls in models for real scalar fields}
\author{D. Bazeia, A.S. In\'acio and L. Losano}
\affiliation{Departamento de F\'\i sica, Universidade Federal da Para\'\i ba,
Caixa Postal 5008, 58051-970 Jo\~ao Pessoa, Para\'\i ba, Brazil}
\date{\today}

\begin{abstract}
We investigate several models described by real scalar fields,
searching for topological defects, and investigating their
linear stability. We also find bosonic zero modes and examine the
thermal corrections at the one-loop level. The classical investigations
are of direct interest to high energy physics and to applications in
condensed matter, in particular to spatially extended systems where fronts
and interfaces separating different phase states may appear. The thermal
investigations show that the finite temperature corrections that appear
in a specific model induce a second order phase transition in the system,
although the thermal effects do not suffice to fully restore the symmetry
at high temperature.
\end{abstract}
\maketitle

\section{Introduction}
\label{intro}

Models described by real scalar fields in $(1,1)$
space-time dimensions are among the simplest systems that support
topological solutions. Usually, these topological solutions are named kinks,
which are classical static solutions of the equations of motion, and the
topological behavior is related to the asymptotic form of the field
configurations, which has to differ in both the positive and negative space
directions. To ensure that the classical solutions have finite energy, one
requires that the asymptotic behavior of the solutions is identified
with minima of the potential that defines the system under consideration,
so in general the potential has to include at least two distinct minima
in order for the system to support topological solutions.

We can investigate real scalar fields in $(3,1)$ space-time dimensions, and
now the topological solutions are named domain walls. These domain walls are
bi-dimensional structures that carry surface tension, which is identified with
the energy of the classical solutions that spring in $(1,1)$ space-time
dimensions. The domain wall structures are supposed to play a role in
applications to several different contexts, ranging from the low energy scale
of condensed matter \cite{esc81,ku84,sle98,wal97} up to the high energy scale
required in the physics of elementary particles, fields and cosmology
\cite{raj82,ktu90,vsh94}.

There are at least two classes of models that support
kinks or domain walls, and we further explore such models in the next
Sec.~{\ref{gen}}. In the first class of models one deals with a single
real scalar field, and the topological solutions are structureless.
Examples of this are the $\phi^4$ and $\phi^6$ models
\cite{raj82}. In the second class of models we deal with systems defined
by two real scalar fields, and now one opens two new possibilities: domain
walls that admit internal structure \cite{mke,mor,brs,97,98,mor98,bbb99},
and junctions of domain walls, which appear in models of two fields when
the potential contains non-collinear minima, as recently investigated for
instance in Refs.{\cite{99a,99b,99c,00a,00b,00c,agm,h,00d,bv01,bb01,n,nns01}}.

There are other motivations to investigate
domain walls in models of field theory, one of them being related to
the fact that the low energy world volume dynamics of branes in string and
M theory may be described by standard models in field
theory \cite{s1,s2,s3}. Besides, one knows that field theory models
of scalar fields may also be used to investigate properties of quasi-linear
polymeric chains, as for instance in the applications of
Refs.~{\cite{96,99,00,01}}, to describe solitary waves in ferroelectric
crystals, the presence of twistons in polyethylene,
and solitons in Langmuir films.

Domain walls have been observed in several
different scenarios in condensed matter, for instance in ferroelectric
crystals \cite{sle98}, in one-dimensional nonlinear lattices \cite{prl},
and more recently in higher spatial dimensions -- see \cite{nat} and references
therein. The potentials that appear in the models of field theory that we
investigate in the present work are also of interest to map systems described
by the Ginzburg-Landau equation, since they may be used to explore the
presence of fronts and interfaces that directly contribute to pattern
formation in reaction-diffusion and in other spatially extended, periodically
forced systems \cite{ku84,wal97,cou90,ce90,ehm98,be98}.

In the present work, in Sec.~{\ref{gen}} we review some known facts
on kinks and domain walls, and there we also introduce other results.
In particular, we present some new models described by one and by two real
scalar fields. The investigations follow in Sec.~{\ref{cla}} where we
search for the topological structures that generate kinks and walls. We
reserve for Sec.~{\ref{stab}} the study of stability of the solutions
that we found in the former section. We investigate the finite temperature
effects in Sec.~{\ref{quan}}, where we examine the effective potential
for a specific model, which engenders three minima at the classical level.
We end the work in Sec.~{\ref{con}}, where we comment on
applications to nonlinear science and briefly review the results
of the present investigation.

\section{General Considerations}
\label{gen}

In this work we are interested in field theory models that describe real
scalar fields and support topological solutions of the
Bogomol'nyi-Prasad-Sommerfield (BPS) type \cite{b,ps}.
In the case of a single real scalar field $\phi$, we consider
the Lagrange density
\be
\label{1f}
{\cal L}=\frac12\partial_{\alpha}\phi\partial^{\alpha}\phi-V(\phi)
\ee
Here $V(\phi)$ is the potential, which identifies the particular model
under consideration. We write the potential in the form
$V(\phi)=(1/2)\,W^2_{\phi}$,
where $W=W(\phi)$ is a smooth function of the field
$\phi$, and $W_\phi=dW/d\phi$. In a supersymmetric theory $W$ is the
superpotential, and this is the way we name $W$ in this work.

The equation of motion for $\phi=\phi(x,t)$ has the general form
\be
\label{em2}
\frac{\partial^2\phi}{\partial t^2}-
\frac{\partial^2\phi}{\partial x^2}+\frac{dV}{d\phi}=0
\ee
and for static solutions we get
\be
\label{ems}
\frac{d^2\phi}{dx^2}=W_{\phi}W_{\phi\phi}
\ee
It was recently shown in Refs.~{\cite{bms01a,bms01b}} that this equation
of motion is equivalent to the first order equations
\be
\label{emf}
\frac{d\phi}{dx}=\pm W_{\phi}
\ee
if one is searching for solutions that obey the boundary conditions
$\lim_{x\to-\infty}\phi(x)={\bar\phi}_i$ and
$\lim_{x\to-\infty}(d\phi/dx)=0$, where ${\bar\phi}_i$ is one among the
several vacua $\{{\bar\phi}_1,{\bar\phi}_2,...\}$ of the system.
In this case the topological solutions are BPS (+) and anti-BPS (-)
solutions. Their energies get minimized to the value $t^{ij}=|\Delta W_{ij}|$,
where $\Delta W_{ij}=W_i-W_j$, with $W_i$ standing for $W({\bar\phi}_i)$.
The BPS and anti-BPS solutions are defined by two vacuum states belonging
to the set of minima that identify the several topological sectors of
the model.

In the case of two real scalar fields $\phi$ and $\chi$ the potential
is written in terms of the superpotential, in a way such that
$V(\phi,\chi)=(1/2)\,W^{2}_{\phi}+(1/2)\,W^{2}_{\chi}$.
The equations of motion for static fields are
\be
\frac{d^{2}\phi}{dx^{2}}=W_{\phi}W_{\phi\phi}+W_{\chi}W_{\chi\phi}
\ee
\be
\frac{d^{2}\chi}{dx^{2}}=W_{\phi}W_{\phi\chi}+W_{\chi}W_{\chi\chi}
\ee
which are solved by the first order equations
\be
\frac{d\phi}{dx}=W_{\phi}\,\;\;\;\;\;\;\;\;
\frac{d\chi}{dx}=W_{\chi}
\ee
or
\be
\frac{d\phi}{dx}=-W_{\phi}\,\;\;\;\;\;\;\;\;
\frac{d\chi}{dx}=-W_{\chi}
\ee
Solutions to these first order equations are BPS (+) and anti-BPS (-)
states. They solve the equations of motion, and have energy minimized to
$t^{ij}=|\Delta W^{ij}|$ as in the case
of a single field; here, however,
$\Delta W^{ij}=W(\phi_i,\chi_i)-W(\phi_j,\chi_j)$, since now we need
a pair of numbers $(\phi_i,\chi_i)$ to represent each one of
the vacuum states in the system of two fields. In the plane
$(\phi,\chi)$ we may have minima that are non collinear, opening the
possibility for junctions of defects. In the case of two real scalar fields,
we can find a family of first order equations that are equivalent to the pair
of second order equations of motion, but this requires that
$W_{\phi\phi}+W_{\chi\chi}=0$, in the case of harmonic superpotentials
\cite{bms01a,bms01b}.

We now turn attention to kinks and domain walls. Perhaps the most
known example of this is given by the $\phi^4$
model, defined by the potential $V(\phi)=(1/2)\,(\phi^2-1)^2$.
Here we are using natural units, and dimensionless fields and coordinates.
In this model the domain wall can be represented by the solution
$\phi_s(x)=\pm\tanh(x)$. The above potential can be written with the
superpotential $W(\phi)=\phi-\phi^3/3$, and the domain wall is of the
BPS or anti-BPS type. The wall tension corresponding to the BPS wall
is $t_s=4/3$.

We can also find structureless domain walls in other models, for instance
in the $\phi^6$ model, which is described by the potential
$V(\phi)=(1/2)\,\phi^2\,(\phi^2-1)^2$.
Here we have
$W(\phi)=(1/2)\phi^2-(1/4)\phi^4$, and the wall configurations are also of the
BPS type, and are given by ${\bar\phi}^2_s=(1/2)[1\pm\tanh(x)]$.
The wall tension is now ${\bar t}_s=1/4$. This potential was investigated
for instance in Ref.~{\cite{loh79}}.

We can build another class of models, which is described
by two real scalar fields. In this case the domain
walls may engender internal structure. This line of investigation follows
as in Refs.~{\cite{mke,mor,brs}} and we illustrate such possibility with
the system defined by the potential
\ben
\label{p2f}
V(\phi,\chi)&=&\frac12(\phi^2-1)^2+
\frac12r^2\left(\chi^2-\frac1r\right)^2+\nonumber\\
& &r(1+2r)\phi^2\chi^2
\een
where the parameter $r\neq0$ is real. This model was first investigated
in Ref.~\cite{95}. This potential follows from the superpotential
$W(\phi,\chi)=\phi-(1/3)\phi^3-r\phi\chi^2$, and the system supports the
two-field solutions $\phi(x)=\tanh(2rx)$ and $\chi(x)=a\,{\rm sech}(2rx)$,
with $a^2=1/r-2$. These solutions are BPS solutions, and now the parameter
$r$ is restricted to the interval $r\in(0,1/2)$. The limit
$r\to1/2$ lead us to the one-field solution
$\phi(x)=\tanh(x)$ and $\chi(x)=0$. The two-field solutions obey
$\phi^2+\chi^2/a^2=1$, which describes an elliptic arc connecting
the two minima $(\pm1,0)$ of the corresponding potential in the
$(\phi,\chi)$ plane. The one-field solutions represent standard
domain walls, while the two-field solutions may be seen as domain
walls having internal structure: the vector $(\phi,\chi)$ in configuration
space describes an straight line segment for the one-field solution, and
an elliptic arc for the two-field solution, resembling light in the
linearly and elliptically polarized cases, respectively. The same
solutions appear in condensed matter, in the anisotropic $XY$ model used
to describe ferromagnetic transition in magnetic systems, and there they
are named Ising and Bloch walls, respectively -- see for instance
Ref.~{\cite{wal97}}, chapter 7.

In the present work we are especially interested in three other models.
They are defined by potentials that describe a single real scalar field
\ben
\label{v1}
V_1(\phi)&=&\frac12-|\phi|+\frac12\phi^2
\\
\label{v2}
V_2(\phi)&=&\frac12\phi^2-|\phi|\,\phi^2+\frac12\phi^4
\een
and two real scalar fields
\ben
\label{v3}
V_3(\phi,\chi)&=&\frac12(1-|\phi|)^2-r(1-|\phi|)\chi^2+\nonumber
\\
& &\frac12r^2\chi^4+2r^2\phi^2\chi^2
\een
The first model was investigated in Refs.~{\cite{hkd77,tdl79}} to model
an exactly soluble linear chain system. It has also been investigated more
recently in Ref.~{\cite{the99} and also in Ref.~{\cite{abl01}}, with
different motivations. The other models are new, and some of their
classical and quantum features will be examined below. 

We notice that the above models can be described by the
following superpotentials
\ben
\label{w1}
W_1(\phi)&=&\phi-\frac12\,|\phi|\,\phi
\\
\label{w2}
W_2(\phi)&=&\frac12\,\phi^2-\frac13\,|\phi|\,\phi^2
\een
and also
\ben
\label{w3}
W_3(\phi,\chi)&=&\phi-\frac12\,|\phi|\,\phi-r\,\phi\,\chi^2
\een
The presence of superpotentials help simplifying the investigation, since
the BPS solutions satisfy first order differential equations, which are
simpler to solve compared to the equations of motion.

Our interest in the first model is directly related to the fact that it
is very much similar to the standard $\phi^4$ model. Thus, we explore
its classical solutions, to see how similar they are to that of the $\phi^4$
model. The same for the second model, which was invented to be similar to
the $\phi^6$ model, and also for the last model, which is an extension of
the first model to the case of two fields, in a way similar to the model
defined by Eq.~(\ref{p2f}), as done in Ref.~{\cite{95}}. These potentials
are also of interest in condensed matter, in the investigation of spatially
extended, periodically forced systems, governed by the Ginzburg-Landau
equation. Explicitly, the Ginzburg-Landau equation for a single real
order parameter $A=A(x,t)$ can be written in the form
\be\label{gl}
\frac{\partial A}{\partial t}=\frac{\partial^2A}{\partial x^2}-\frac{dU}{dA}
\ee
where $U=U(A)$ is a function of $A$, which may be identified with the
potential of the field models that we are using in the present work. These
models may then provide diverse scenarios for fronts and interfaces to
travel according to the Ginzburg-Landau dynamics. 

\section{Topological solutions}
\label{cla}

In this section we search for BPS solutions in the models introduced
in the former Sec.~{\ref{gen}}. We investigate the first order equations
corresponding to each one of the three models separately.

\subsection{Model 1}
\label{m1}

This model is defined by the superpotential of Eq.~(\ref{w1}).
In this case there are two singular points, that represents the two minima
of the potential in Eq.~(\ref{v1}). These minima are ${v_{1}=1, v_{2}=-1}$.
The equation of motion for ${\phi=\phi(x)}$ is
\be
\frac{d^2\phi}{dx^2}=\phi - \frac{\phi}{|\phi|}
\ee
We are searching for topological solutions, so we can consider instead of
the second order differential equation the two first order equations
\be
\frac{d\phi}{dx}=\pm(1-|\phi|)
\ee
They support the BPS states
\be
\label{sm1}
\phi_\pm(x)=\pm 2\frac{\tanh(x/2)}{1+\tanh(|x|/2)}
\ee
After inspecting the explicit form of these BPS solutions, we notice
that they have the standard kink-like profile, as in the $\phi^4$ model,
as we show in Fig.~1. Also, they are linearly stable, minimizing the energy
to $t_1=1$, below the value $t_s=4/3$ that appears in the $\phi^4$ model.

The interesting feature of this model is that at the classical level it
is similar to the $\phi^4$ model, although the potential
goes up to the second order power in the field. We illustrate the present
model and its topological solutions in Fig.~1, where we also plot the
topological state of the standard $\phi^4$ model to offer a visual
comparison between the BPS states of the two models. There we notice that
the BPS states of model 1 are thicker than the corresponding states of the
$\phi^4$ model. Also, we notice that the tension of these BPS states
can be written as $t_1=(3/4)t_s$.

We notice that this model may be used as an alternative to the $\phi^4$ model,
in applications in condensed matter. For instance, it may represent the
motion in the polyacetylene (PA) chain, describing single-double
or double-single bound alternation in the PA chain, but this is out
of the scope of the present work.

\subsection{Model 2}
\label{m2}

Now we explore the second model, described by the potential of Eq.~(\ref{v2}).
It can be written in terms of the superpotential of Eq.~(\ref{w2}).
This model contains three minima, at the values ${v_{1}=1,v_{2}=-1,v_{3}=0}$.
It supports two topological sectors of the BPS type. These
sectors are degenerate, and the wall tension is given by $t_2=1/6$. 
The BPS solutions satisfy
\be
\frac{d\phi}{dx}=\pm\phi(1-|\phi|)
\ee
The solutions are
\ben
\label{sm2+}
\phi^{+}_{\pm}(x)=\pm\frac12[1+\tanh({x/2})]
\\
\label{sm2-}
\phi^{-}_{\pm}(x)=\pm\frac12[1-\tanh({x/2})]
\een
They are similar to the kink solutions that one finds in the $\phi^6$ model,
as we have shown in Sec.~{\ref{gen}}. We can write the wall tension in
this second model as $t_2=(2/3){\bar t}_s$, where ${\bar t}_s$ is the
tension value for the BPS solutions of the $\phi^6$ model. We
illustrate the present model and its topological solutions in Fig.~2.
The interesting feature of the second model is that its potential is
polynomial, of the fourth order type, but it presents two distinct phases,
the asymmetric phase and another one, symmetric, represented by the vanishing
minimum, $v_3=0$.

%%%%%%%%%%%%%%%%%%%%%%%%%%%%%%%%%%%%%%%%%%%%%%%%%%%%%%%%%%%%%%%%%%%%%%%%%%
\begin{figure}[h]
\includegraphics[{height=5cm,width=8cm}]{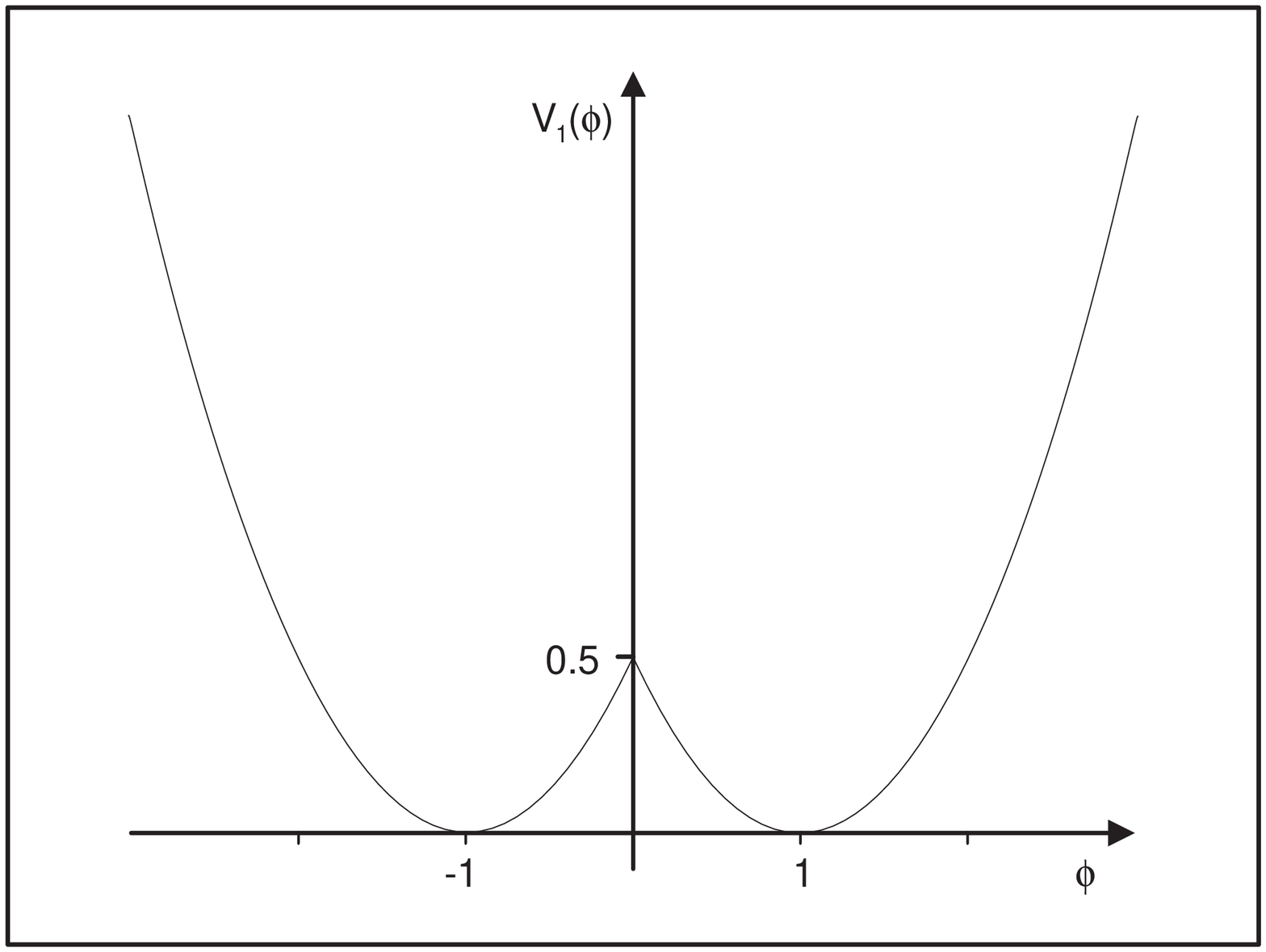}
\includegraphics[{height=5cm,width=8cm}]{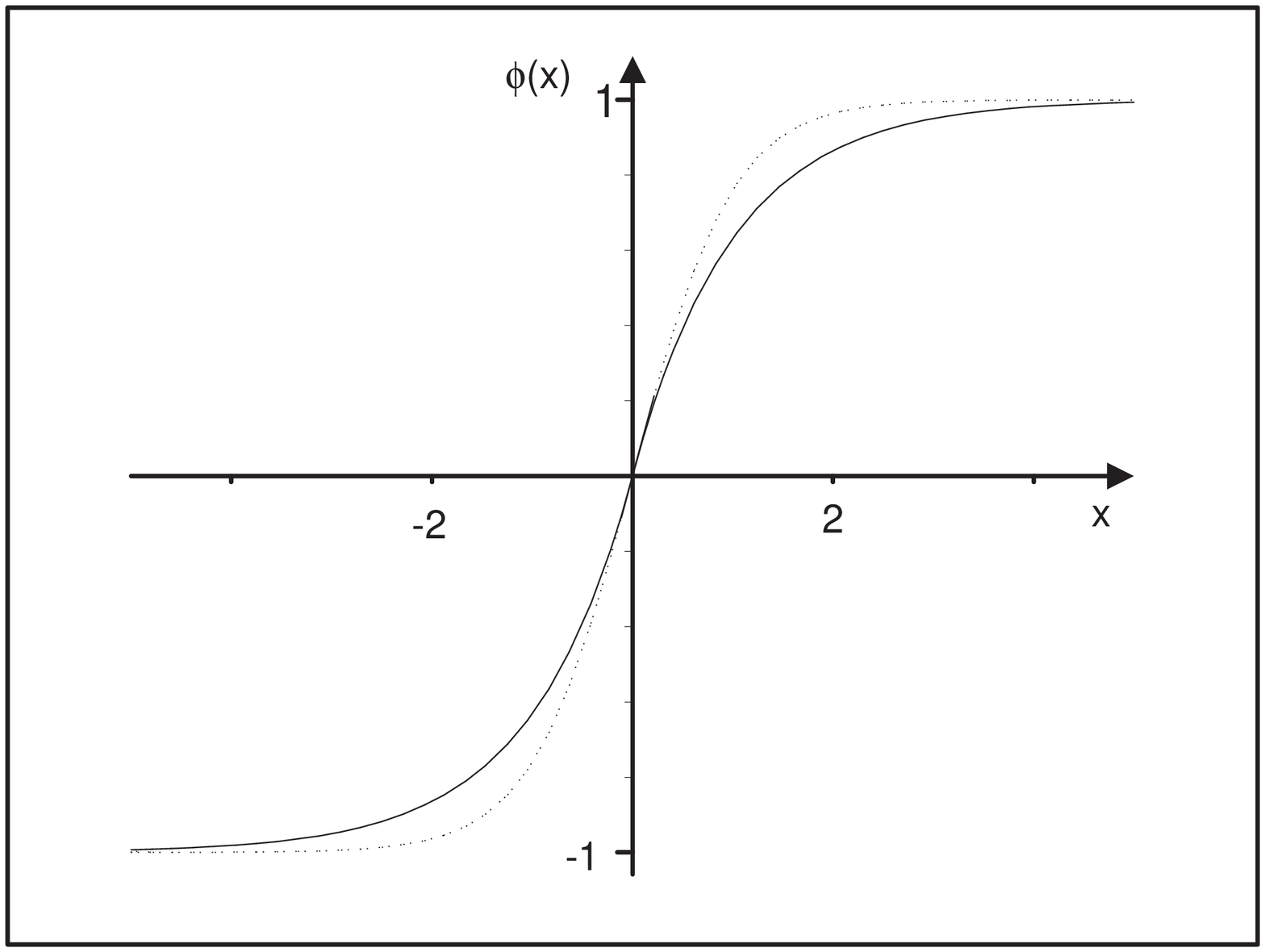}
\vspace{0.3cm}
\caption{The potential $V_1(\phi)$ (right) given in Eq.(\ref{v1}),
and the respective topological solution (left). For comparison, the dotted
line shows the topologial solution of the $\phi^4$ model. All quantities
that appear in these figures are dimensionless.}
\end{figure}
%%%%%%%%%%%%%%%%%%%%%%%%%%%%%%%%%%%%%%%%%%%%%%%%%%%%%%%%%%%%%%%%%%%%%%%%%%

The second model seems appropriate to simulate a first order phase
transition. However, differently from the $\phi^6$ system the present model
has the property of having the same squared mass, irrespective of the
symmetric or asymmetric phases. To see this we use the potential
$(\ref{v2})$ to obtain
\be
\frac{d^2V_2}{d\phi^2}=1-6\,|\phi|+6\,\phi^2
\ee
so that $m_2^2(\phi=0)=m_2^2(\phi=\pm1)=1$. As we shall see below,
this fact also shows that $U(x)=d^2V_2/d\phi^2$ is symmetric respect to $x$
when calculated at any BPS state, so that the investigation to find the first
quantum corrections to the energy of the BPS solutions follows naturally,
cincumventing the intricacy that happens to appear in the standard $\phi^6$
model \cite{loh79}. The form of the potential of Eq.~(\ref{v2})
shows a symmetry respect to $\phi=\pm1/2$ for $\phi$ in the interval
$0\leq\phi\leq1$ -- see Fig.~2. This feature does not appear
in the $\phi^6$ model.

The classical picture that appears for the second model is somehow similar
to the picture of the $\phi^6$ model. Thus, it may be used as a new model
to explore conformational properties of polyethylene (PE), as an alternative
to the case investigated in Refs.~{\cite{99,00}}. Moreover, since its potential
contains up to the fourth order power in the field, in Sec.~{\ref{quan}}
we investigate the semiclassical or one-loop corrections, to see how the
thermal effects change the classical picture.

%%%%%%%%%%%%%%%%%%%%%%%%%%%%%%%%%%%%%%%%%%%%%%%%%%%%%%%%%%%%%%%%%%%%%%%%%%
\begin{figure}[h]
\includegraphics[{height=5cm,width=8cm}]{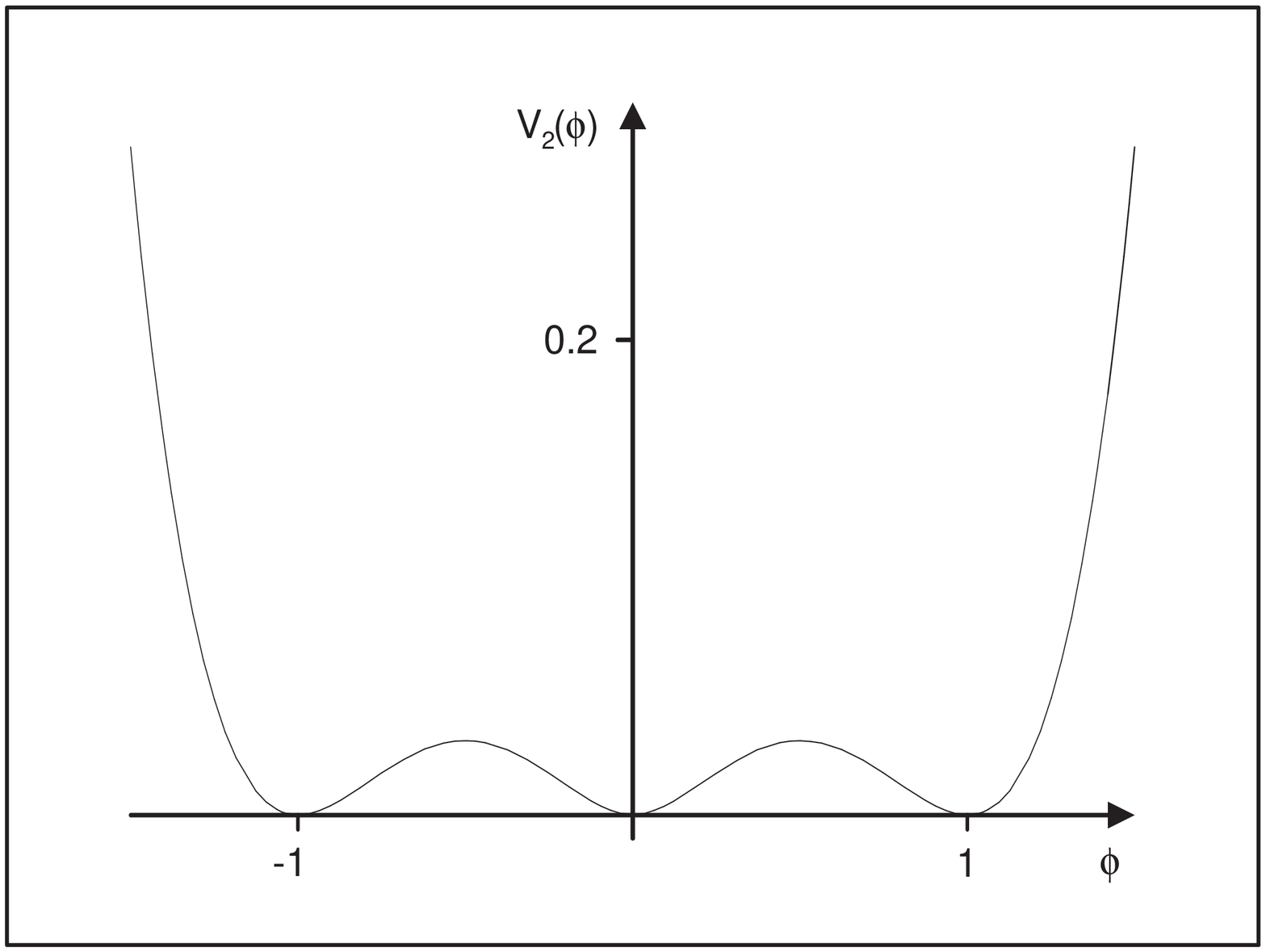}
\includegraphics[{height=5cm,width=8cm}]{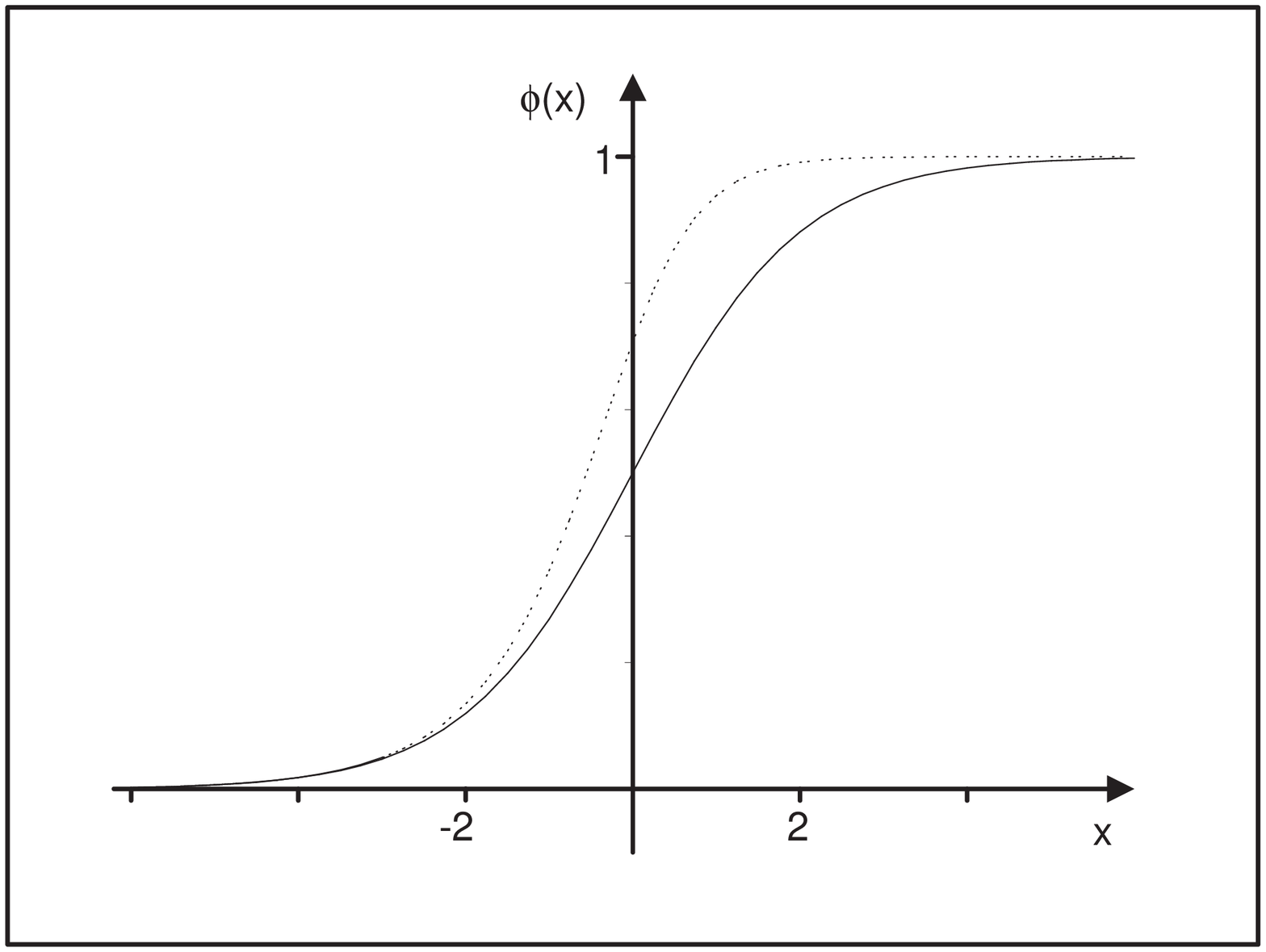}
\vspace{0.3cm}
\caption{The potential $V_2(\phi)$ (right) given in Eq.(\ref{v1}),
and the respective topological solution (left). For comparison, the dotted line
shows the kink of the $\phi^6$ model, which is asymmetric. All quantities
that appear in these figures are dimensionless.}
\end{figure}
%%%%%%%%%%%%%%%%%%%%%%%%%%%%%%%%%%%%%%%%%%%%%%%%%%%%%%%%%%%%%%%%%%%%%%%%%%

\subsection{Model 3}

The third model is given by the potential of Eq.~(\ref{v3}).
In terms of the superpotential of Eq.~(\ref{w3}) the first order equations
are
\ben
\frac{d\phi}{dx}&=&1-|\phi|-r\chi^{2}
\\
\frac{d\chi}{dx}&=&-2r\phi\chi 
\een
There are four degenerate minima for $r>0$, at the points
$v_{1,2}=(\pm1,0)$ and $v_{3,4}=(0,\pm1/\sqrt{r})$. There are five BPS
sectors, and only one non-BPS sector, which is the sector connecting 
the minima $v_{3}$ and $v_{4}$. The tensions of the BPS states are
$t_3^{1}= 1$ and $t_3^{2}=\sqrt{r}$.

This model is similar to the model considered in Eq.~(\ref{p2f}).
To find BPS solutions we follow Refs.~{\cite{00b,95}}. In the 
sector connecting the minima $v_{1}$ and $v_{2}$ we have
\be
\label{sm3}
\phi_\pm(x)=\pm 2\frac{\tanh(x/2)}{1+\tanh(|x|/2)},\;\;\;\;\chi(x)=0
\ee
The corresponding orbit is a straight line segment
joinning the minima $v_{1}$ and $v_2$. We have been unable
to find non trivial analytical BPS states in this sector,
for solutions describing an elliptic orbit, although they
appear in the model of Eq.~(\ref{p2f}).

There are other BPS states, in the sectors that connect the minimum $v_1$
or $v_2$ to $v_3$ or $v_4$. There are several cases, and we get explicit
solutions for the specific value $r=1/2$, which form straight line segments
connecting $v_1$ or $v_2$ to $v_3$ or $v_4$. The orbits are described by
configurations such that $\chi=\pm\sqrt{2}(\phi\mp1)$. The explicit solutions
are, in the sectors connecting $v_2$ to $v_3$ or $v_4$
\be\label{m311a}
\phi(x)=-\frac12[1-\tanh(\frac{x}{2})]
\ee
\be\label{m312a}
\chi(x)=\pm\frac{\sqrt{2}}{2}[1+\tanh(\frac{x}{2})]
\ee
and in the sectors connecting $v_{1}$ to $v_{3}$ or $v_{4}$
\be
\label{m311}
\phi(x)=\frac12[1+\tanh(\frac{x}{2})]
\ee
\be
\label{m312}
\chi(x)=\pm\frac{\sqrt{2}}{2}[1-\tanh(\frac{x}{2})]
\ee
These solutions are similar to the solutions found in Ref.~{\cite{shi98}},
for the model of Eq.~(\ref{p2f}). In Fig.~3 we plot one of the several
BPS states that appear in this case.

The nontrivial BPS solutions that we found in model 3 can be used to model
domain walls solutions that appear in models for binary mixtures
of Bose-Einstein condensates (BEC) \cite{mya}. As one knows, for different
coupling coefficient between the two BECs, the mixture exhibit complex
spatial structure which may be described by domain walls \cite{tim}. More
importantly, in Ref.~{\cite{cha}} one has realized that these domain
walls consist of multicomponent solitons. And these multicomponent walls
are very much similar to the two-field solutions that we have just found
in model 3. A specific feature of the BPS state shown in Fig.~3 is that
the limit $x\to0$ gives $\phi(0)>\chi(0)$, leading to unequal components
in the binary mixture. We shall further explore this issue elsewhere. 

This model is also of interest within the context of of spatially extended
systems, where phase fronts separating different phase states may appear.
The two field model that we are investigating may be used to describe a
complex order parameter $A=\phi+i\chi$, ${\bar A}=\phi-i\chi$ which obeys
the Ginzburg-Landau equation
\be\label{glc}
\frac{\partial A}{\partial t}=\frac{\partial^2A}{\partial x^2}-
\frac{\partial V}{\partial{\bar A}}
\ee
where $V=V({\bar A}A)$ is the potential of Eq.~(\ref{v3}), described in terms
of the complex order parameter. This equation generalizes Eq.~(\ref{gl})
to the case of a complex order parameter, and the solutions that we have
found in Eqs.~(\ref{sm3}), and (\ref{m311a}) and (\ref{m312a}) or
(\ref{m311}) and (\ref{m312}) can be used to describe $\pi$, and $\pi/2$
fronts, respectively \cite{ehm98}.

%%%%%%%%%%%%%%%%%%%%%%%%%%%%%%%%%%%%%%%%%%%%%%%%%%%%%%%%%%%%%%%%%%%%%%%%%%
\begin{figure}[h]
\includegraphics[{height=5cm,width=10cm}]{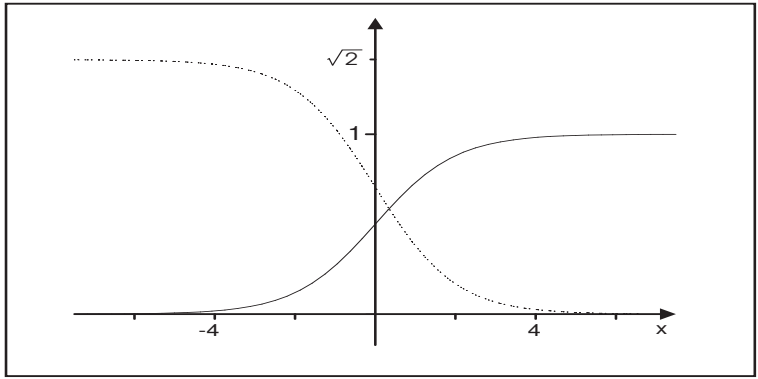}
\vspace{0.3cm}
\caption{A tipical solution that appear from Eqs.~(\ref{m311}) (thin line)
and (\ref{m312}) (dotted line), to illustrate how the fields behave as a
nontrivial BPS state. All quantities that appear in this figure
are dimensionless.}
\end{figure}
%%%%%%%%%%%%%%%%%%%%%%%%%%%%%%%%%%%%%%%%%%%%%%%%%%%%%%%%%%%%%%%%%%%%%%%%%%

%%%%%%%%%%%%%%%%%%%%%%%%%%%%%%%%%%%%%%%%%%%%%%
\section{Stability}
\label{stab}

The study of linear stability of BPS solutions is of direct interest to high
energy physics. The point here is that an investigation at the quantum level
would require the explicit calculation of eigenvalues and eigenfunctions that
appear from linear stability. To illustrate the subject, let us    
investigate linear stability of the BPS solutions obtained in the former
Sec.~{\ref{cla}}. In the case of a single field the investigation
requires that one uses
\be
\label{sta}
\phi(x,t)=\phi_{s}(x)+\sum_{n}\,\eta_{n}(x)\,\cos(w_{n}t).
\ee
where $\phi_s(x)$ stands for the static field, and the remaining terms
represent the fluctuations about it. Linear stability implies that the
fluctuations remains limited in time, so that the set of frequencies
$\{w_n\}$ forms a set of real numbers.

We use Eq.~(\ref{sta}) into the second order equation of motion (\ref{em2})
to obtain the Schr\"odinger-like equation $H\eta_n(x)=w_n^2\,\eta_n(x)$,
where
\be
H=-\frac{d^{2}}{dx^{2}}+ U(x)
\ee
and
\be
U(x)=W^{2}_{\phi\phi}- W_{\phi}W_{\phi\phi\phi}.
\ee
which must be calculated with the static configuration, the BPS states.

As we know, we end up with a supersymmetric quantum mechanical problem. Thus,
we generalize the above situation by introducing the pair of Hamiltonians
\be
H_{\pm}=-\frac{d^{2}}{dx^{2}}+ U_{\pm}(x)
\ee
where
\be
U_{\pm}(x)=W^{2}_{\phi\phi}\mp W_{\phi}W_{\phi\phi\phi}.
\ee
These Hamiltonians can be factorized as
\be
H_{\pm}=A^{\dag}_{\pm}\,A_{\pm}
\ee
where $A_{\pm}$ are first order operators, given by
\be
A_{\pm}=-\frac{d}{dx}\pm W_{\phi\phi}
\ee
This factorization ensures that the Hamiltonians $H_{\pm}$ are positive
definite, so that their eigenvalues are all non-negative real numbers,
thus ensuring linear stability of the classical topologocal solutions.

\subsection{Linear stability: model 1}
\label{lsm1}

In order to investigate linear stability of the BPS states obtained
in the former Sec.~{\ref{m1}}, we use Eqs.~(\ref{em2}), (\ref{sm1}),
and (\ref{sta}) to obtain the Schr\"odinger-like equation
\be
\label{sch}
\left(-\frac{d^{2}}{dx^{2}}+1-2\,\delta(x)\right)\eta_{n}(x)=
w_{n}^{2}\eta_{n}(x)
\ee
In this case the potential is given by
\be
U(x)=1-2\delta(x)
\ee

Stability of the classical solution (\ref{sm1}) implies that the eigenvalues
of the above Schr\"odinger-like Eq.~(\ref{sch}) should be non-negative.
But this is indeed the case, because the atractive potential $\delta(x)$
has only one bound state, and the specific form of the potential
$U(x)=1-2\delta(x)$ places this bound state at zero energy. To see this
explicitly we follow the steps already presented and we factorize the
Hamiltonian in the form
\be
-\frac{d^2}{dx^2}+1-2\delta(x)=\left(-\frac{d}{dx}+\frac{|x|}{x}\right)
\left(\frac{d}{dx}+\frac{|x|}{x}\right)
\ee
The single bound state is the zero mode, $\eta_0(x)$, which obeys
\be
\left(\frac{d}{dx}+\frac{|x|}{x}\right)\eta_0(x)=0
\ee
This equation is easily solved to give the normalized zero mode
\be
\eta_0(x)=e^{-|x|}
\ee

The quantum mechanical problem is implemented with the partner Hamiltonians.
They are
\be
H_{\pm}=-\frac{d^2}{dx^2}+U_{\pm}(x)
\ee
where $U_{\pm}(x)=1\pm\,2\,\delta(x)$.
They are obtained via $H_{\pm}=A^{\dag}_{\pm}A_{\pm}$, with the first order
differential operators
\be
A_{\pm}=\pm\frac{d}{dx}+\frac{|x|}{x}
\ee

The bosonic model under investigation can be seen as the bosonic portion
of a more general theory, which includes fermions via the standard Yukawa
coupling. In the model with fermions we can search for fermionic zero modes.
The investigation may follow the lines of Ref.~{\cite{baz99}}.

\subsection{Linear stability: model 2}

We follow the same steps to investigate stability of the BPS
solutions of the second model, given by Eqs.~(\ref{sm2+}) and (\ref{sm2-}).
In this case the Schr\"odinger-like equation becomes
\be
\label{sem2}
\left(-\frac{d^{2}}{dx^{2}}+1-\frac{3}{2}\;
{\rm sech}^{2}\left(\frac{x}{2}\right)\right)\eta_{n}(x)
=w_{n}^{2}\eta_{n}(x)
\ee
This new Hamiltonian can be fatorized as
\be
\left(\frac{d}{dx}-\tanh\left(\frac{x}{2}\right)\right)
\left(-\frac{d}{dx}-\tanh\left(\frac{x}{2}\right)\right)
\ee
The zero mode obeys
\be
\left(-\frac{d}{dx}-\tanh\left(\frac{x}{2}\right)\right)\eta_0(x)=0
\ee
This first order equation can be integrated easily; the normalized zero mode
has the explicit form
\be
\eta_0(x)=\frac12\sqrt\frac{3}{2}\;{\rm sech}^{2}\left(\frac{x}{2}\right)
\ee

We notice that the above Schr\"odinger-like Eq.~(\ref{sem2}) is the equation
that appears in the modified Poschl-Teller problem, so that the investigation
here goes very much like it does in the standard $\phi^4$ model -- see for
instance Ref.~{\cite{raj82}}. As a result, besides the above zero mode the
present problem has another bound state, at energy $w_1=\sqrt{3}/2$ -- recall
that we are using dimensionless units in the present work. Also, the continuum
starts at energy $w=1$. 

The quantum mechanical problem is implemented by the partner Hamiltonians
$H_{\pm}=A^{\dag}_{\pm}A_{\pm}$, where the first order
differential operators have the explicit form
\be
A_{\pm}=\pm\frac{d}{dx}-\tanh\left(\frac{x}{2}\right)
\ee

\subsection{Linear stability: model 3}

We now direct our attention to the third model. We consider tha classical
solution (\ref{sm3}). In this case we consider $\phi(x,t)$ as before,
and $\chi(x,t)$ in the form 
\be
\chi(x,t)=\sum_{n}\,\xi_{n}(x)\,\cos({\ov w}_{n}t)
\ee
The Schr\"odinger-like equation splits into two equations,
one of them being exactly Eq.~(\ref{sch}), and the other is
\be
\left(-\frac{d^{2}}{dx^{2}}+U(x)\right)\xi_{n}(x)
={\ov w}_{n}^{2}\xi_{n}(x)
\ee
where
\be
U(x)=2r\frac{8r-(8r+1)\,{\rm sech}^2(x/2)}{[1+\tanh(|x|/2)]^2}
\ee
In this case the potential has at least one bound state,
the zero mode. From the analytical point of view the problem is somehow
complicated, and we could not find any explicit solutions. However, we could
verify that the number of bound states depends on $r$, and increases
for increasing $r$. As we can see from Eq.~(\ref{w3}), the parameter $r$ is
related to the way the two fields interact. Thus, one sees that
the number of bound states increases when one increases the strength of the
interaction between the two fields. This feature also appears in other models,
and we shall further explore this specific issue in another work.

%%%%%%%%%%%%%%%%%%%%%%%%%%%%%%%%%%%%%%%%%%%%%%%%
\section{Thermal effects}
\label{quan}

Let us now deal with semiclassical effects. However, instead of calculating
the quantum corrections to the energy of the classical solutions
let us investigate the effective potential in the case of a single real
scalar field. There are several distinct but equivalent ways of implementing
the calculations and we choose to follow Refs.~{\cite{j74,dj74}}. We get the
general expression, which is valid at the one-loop level \cite{j74}
\be
\label{1loop}
V^1(\phi)=\frac12\int\frac{d^4k}{(2\pi)^4}\ln[k^2+V''(\phi)]
\ee
where $V''(\phi)=d^2V/d\phi^2$, and the metric is now Euclidian. The thermal
effects are obtained after changing the one loop contribution
of Eq.~(\ref{1loop}) to
\be
\label{v:the}
V_{\beta}^1(\phi)=\frac12\int
\frac{d^3k}{(2\pi)^3}\sum_{n=-\infty}^{\infty}
\ln\Biggl[\left(\frac{2\pi n}{\beta}\right)^2+{\bf k}^2+V''(\phi)\Biggr]
\ee
The temperature enters the game as $\beta=1/T$. We make the sum to obtain
the finite temperature contribution \cite{dj74,b83,a85}
\be
\label{1lT}
V^{1}_{\beta}(\phi)=\frac{1}{\beta}\int\frac{d^3k}{(2\pi)^3}
\ln\left[1-e^{-\beta\sqrt{{\bf k}^2+V''(\phi)}}\right]
\ee
The temperature dependent contribution can be used to investigate how the
thermal effects enter the game.

We investigate the second model at finite temperature. We rewrite
the potential of Eq.~(\ref{v2}) in the form
\be
V_2(\phi)=\frac12\phi^2(m-\lambda|\phi|)^2
\ee
Here we have introduced $m$ and $\lambda$ to parametrize the potential,
restauring the standard notation that appears with $\hbar=c=1$.
The parameters $m$ and $\lambda$ are real, and we consider $m/\lambda>0$
to allow for spontaneous symmetry breaking. We notice that $\phi$ and
$m$ has dimension of energy, while $\lambda$ is dimensionless.

The motivation to study this model is that it has an interesting property,
not seem in the $\phi^6$ model. The above potential goes up to the fourth
order power in the field, and it has minima at the nonzero
values $|{\bar\phi}|=m/\lambda$ and at $\phi=0$; however, the classical
masses corresponding to these minima degenerate to the single (squared)
value $m^2$.

In the hight temperature limit $(T>>m)$ the one loop effective potential
in this case is well approximated by
\be
V_{eff}(\phi)=\frac12\phi^2(m-\lambda|\phi|)^2+\frac{T^2}{24}\,V''(\phi)
\ee
where $V''(\phi)=m^2-6m\lambda|\phi|+6\lambda^2\phi^2$. We recall that
in natural units the temperature has dimension of energy.
The effective potential has the minima
\be
{\bar\phi}_i=\pm\,\frac12\,\frac{m}{\lambda}\left(1\pm\sqrt{1-
\frac{\lambda^2}{m^2}T^2}\right)
\ee 
This result allows introducing the critical temperature $T_{c}=m/\lambda$,
which identifes two distinct behaviors: for $T$ below $T_c$
the effective potential supports four minima, at the above values
of ${\bar\phi}_i,i=1,2,3,4$, but for $T\geq T_c$ there are
only two minima, at the values $\bar\phi_{\pm}=\pm m/2\lambda$.
The behavior of the effective potential is shown
in Fig.~4, where we depict two tipical cases, for $T=(4/5)T_c$ and for $T=T_c$.

We notice that the high temperature effects
are unable to restore the symmetry, which remains broken for
$T\geq T_c=m/\lambda$, which is greater than $m$ in the weak
coupling limit that makes our calculations reliable. Although this result
goes against conventional wisdow, it has already appeared in other models,
as for instance in \cite{dj74,mse79} and in references therein.

We investigate the (squared) mass, which can be obtained from the effective
potential by the usual procedure. It can be written as
\be
m^2(T)= V''({\bar\phi})+\frac1{24}V''''({\bar\phi})\,T^2
\ee
Thus, for temperature lower than $T_c$ the masses at the constant
configurations ${\bar\phi}_i$ degenerate to the single value
\be
m^2(T)=m^2-\lambda^2T^2,\;\;\;\;\;T\leq T_c
\ee
For temperature higher than $T_c$ we get
\be
m^2(T)=-\frac{m^2}{2}+\frac{\lambda^2}{2}T^2,\;\;\;\;\; T\geq T_c
\ee
These results show that the effective
mass decreases to zero for temperature lower than $T_c$,
and it increases from zero for temperature above $T_c$. The critical
temperature identifies the point where the effective mass vanishes,
and for higher temperature there is no symmetry restoration anymore.

%%%%%%%%%%%%%%%%%%%%%%%%%%%%%%%%%%%%%%%%%%%%%%%%%%%%%%%%%%%%%%%%%%%%%%%%%%
\begin{figure}[h]
\includegraphics[{height=5cm,width=8cm}]{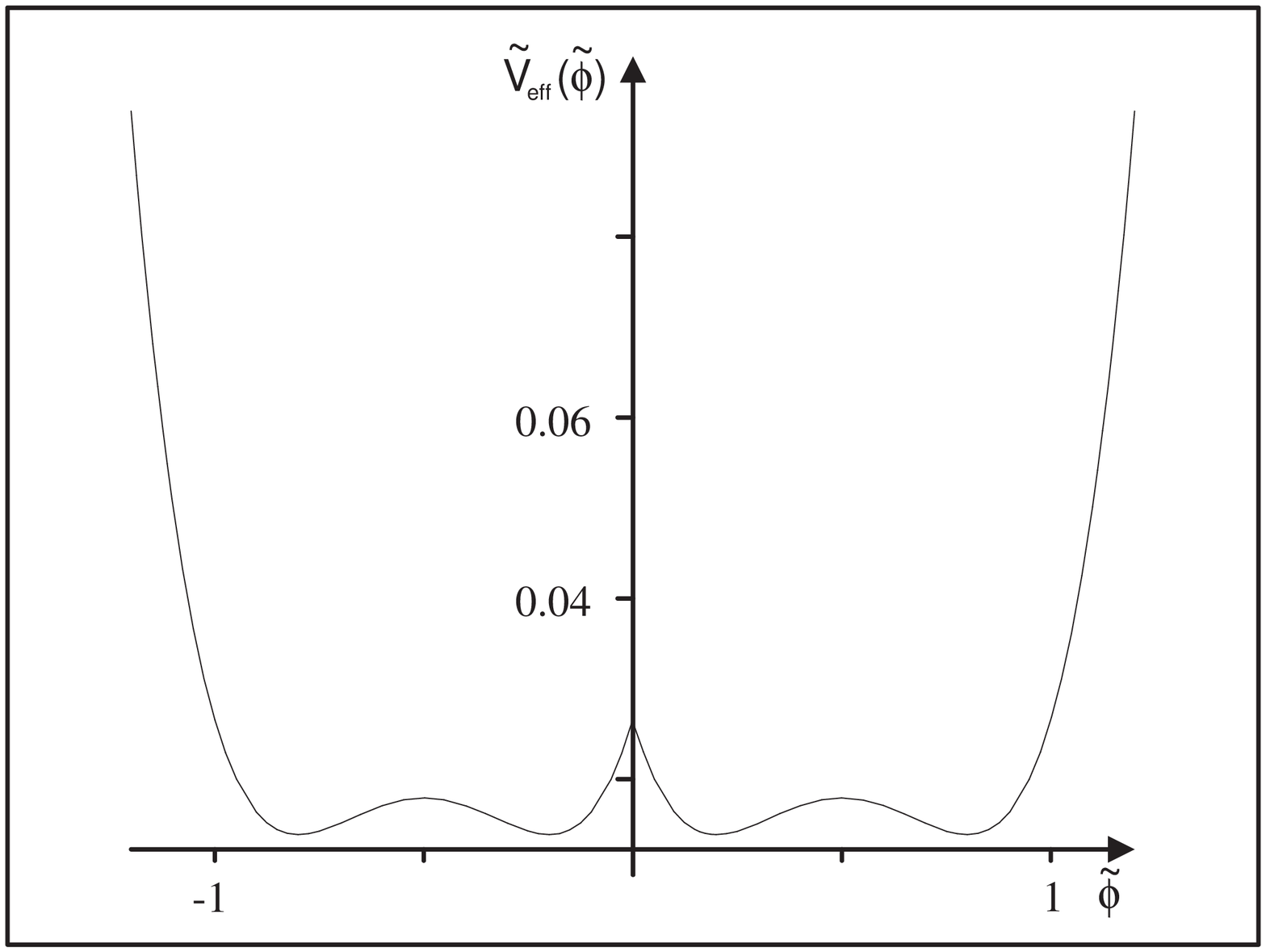}
\includegraphics[{height=5cm,width=8cm}]{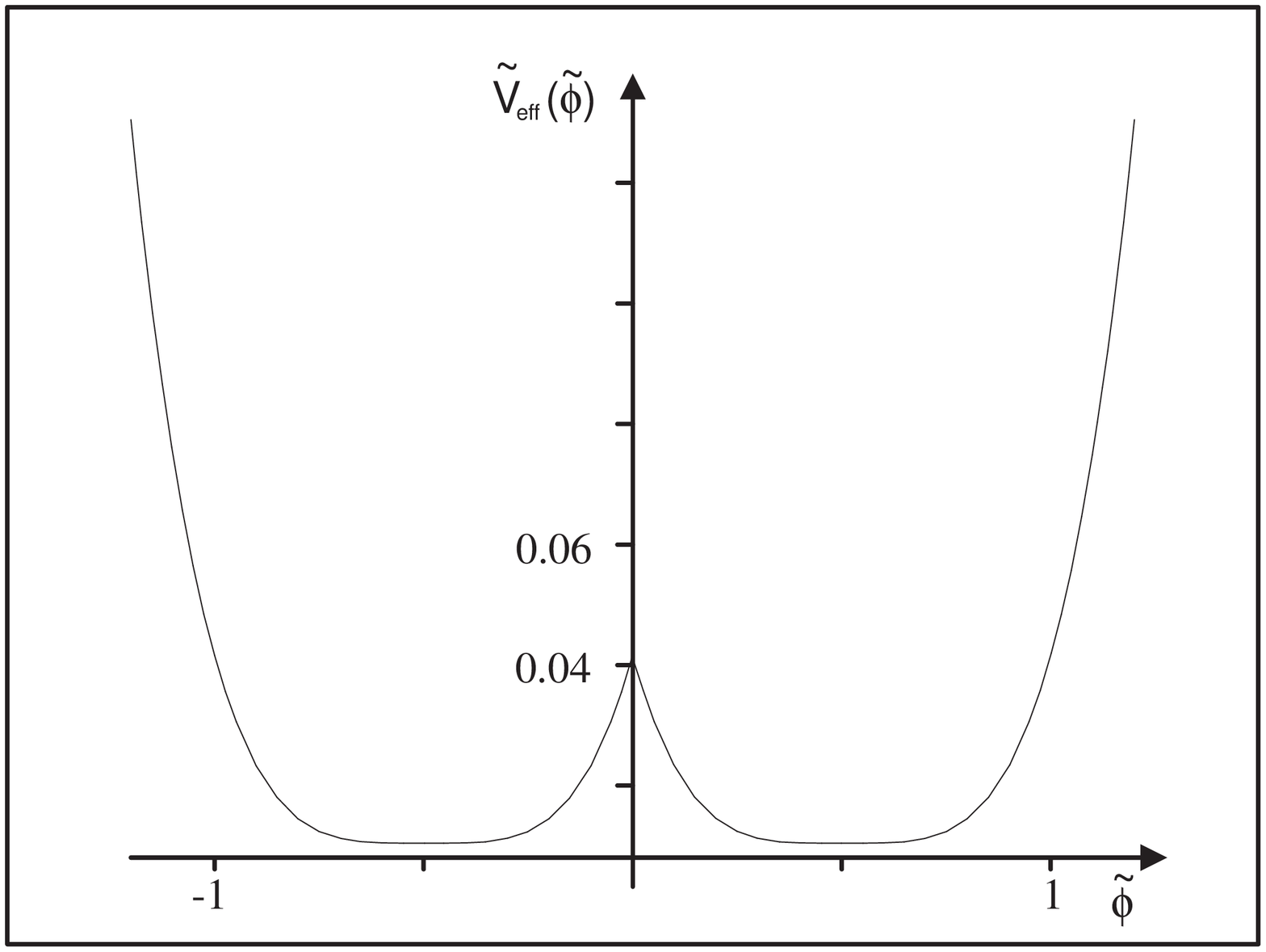}
\vspace{0.3cm}
\caption{The effective potential ${\tilde V}_{eff}({\tilde\phi})$ as a
function of the dimensionless variable ${\tilde\phi}=(\lambda/m)\phi$,
for the two tipical values $T/T_c=4/5$ (right) and $T/T_c=1$ (left).
The effective potential ${\widetilde V}_{eff}$ is dimensionless, given by
${\tilde V}_{eff}({\tilde\phi})=V_{eff}({\tilde\phi})/\lambda^2 T^4_c$.}
\end{figure}
%%%%%%%%%%%%%%%%%%%%%%%%%%%%%%%%%%%%%%%%%%%%%%%%%%%%%%%%%%%%%%%%%%%%%%%%%%

In Fig.~5 we depict the minima of the potential, and the squared mass
at finite temperature. There we see that ${\bar\phi}(T)$ varies continuosly
as the temperature crosses the critical value $T_c=m/\lambda$, indicating the
presence of a second order phase transition.

\section{Conclusions}
\label{con}

In the present work we have investigated several models described by one and
by two real scalar fields. The main investigations concern the search for
BPS states, that is, for topological solutions that solve first order
differential equations. These solutions minimize the energy to the
Bogomol'nyi bound, which is given solely in terms of the superpotential,
and the asymptotic value of the corresponding field configurations.

The search for topological solutions is done at the classical level, and
we have payed special attention to models
introduced in Sec.~{\ref{gen}}, some of them further explored in
Sec.~{\ref{cla}}. These investigations have shown that model 1
and model 2, defined by the potentials of Eqs.~(\ref{v1}) and (\ref{v2}),
respectively, are classically similar to the standard field theory models
known as the $\phi^4$ and the $\phi^6$ models. The other model is a
two-field model, defined by the potential of Eq.~(\ref{v3}). It is 
similar to a model first investigated in Ref.~{\cite{95}}.

%%%%%%%%%%%%%%%%%%%%%%%%%%%%%%%%%%%%%%%%%%%%%%%%%%%%%%%%%%%%%%%%%%%%%%%%%%
\begin{figure}[h]
\includegraphics[{height=5cm,width=8cm}]{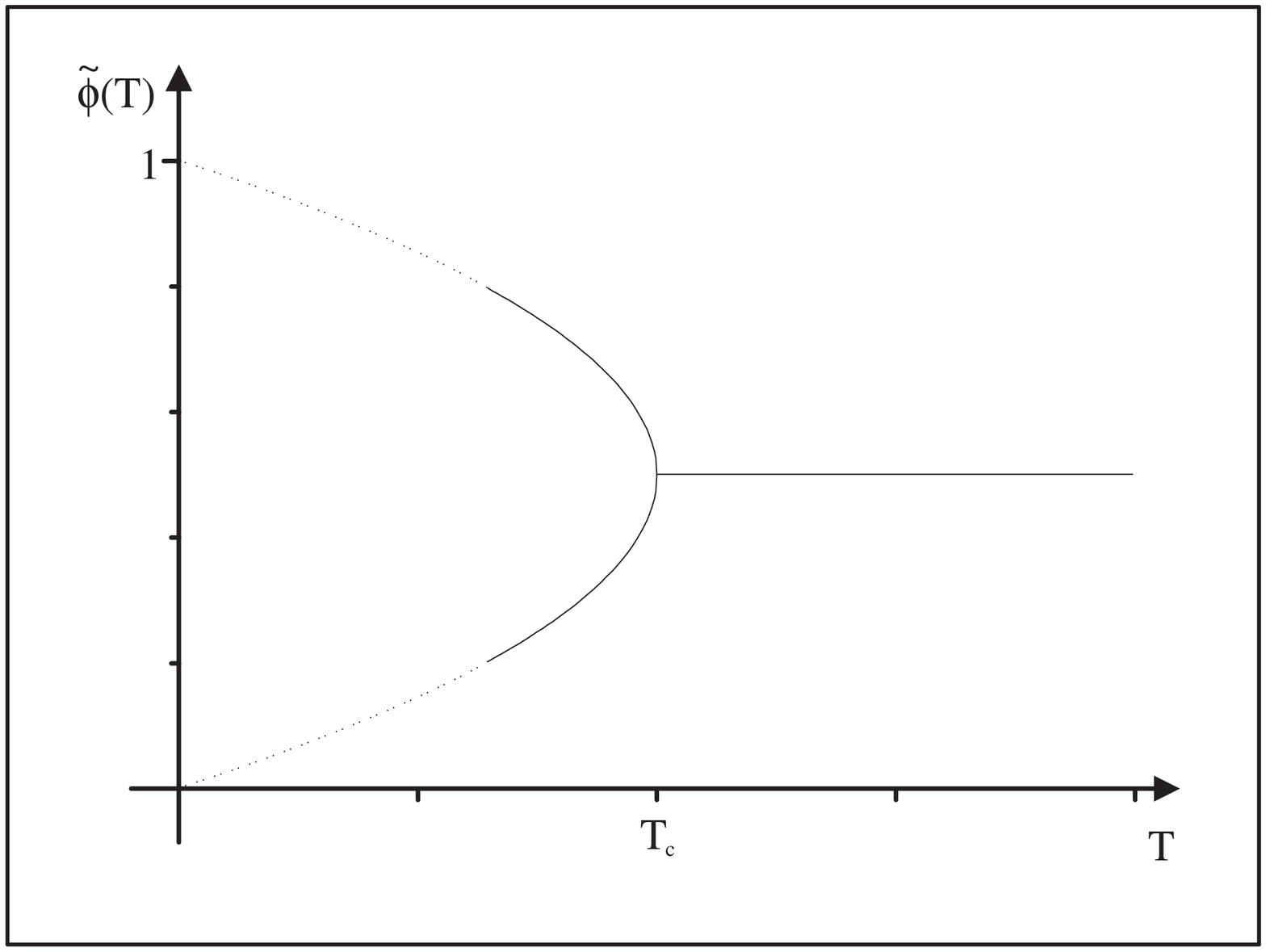}
\includegraphics[{height=5cm,width=8cm}]{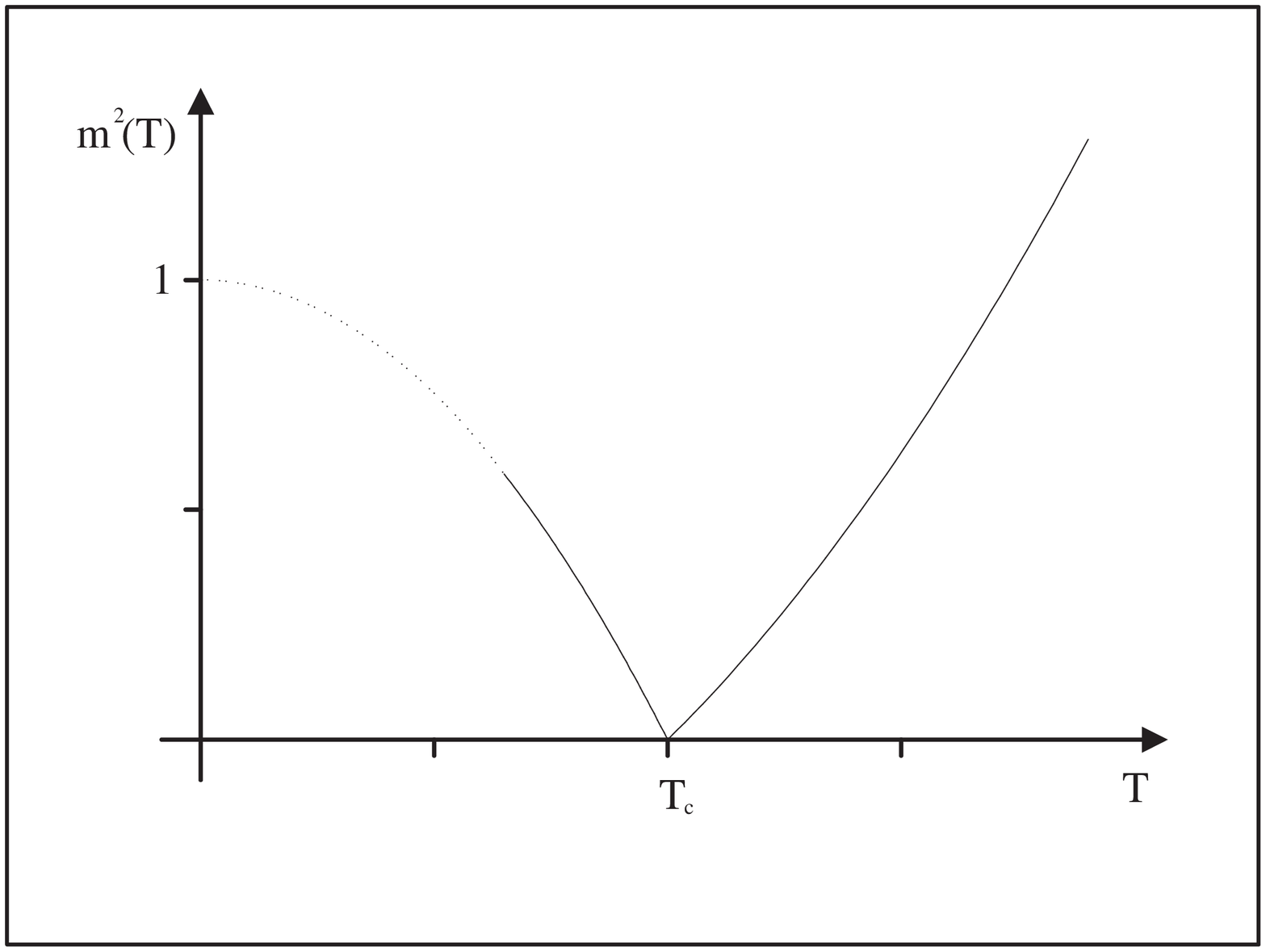}
\vspace{0.3cm}
\caption{The (dimensionless) minima ${\tilde\phi}(T)$
[${\tilde\phi}=(\lambda/m){\bar\phi}$] of the effective
potential (right), and the effective mass $m^2(T)$ (left), depicted as a
function of the temperature. Here we are using natural units, thus the
mass $m$ and the temperature $T$ have dimension of energy, and are measured
in electron-Volt. The dotted-line regions in both figures are depicted to
inform that our results are obtained within the high temperature approximation,
and they may not be valid in low temperatures.}
\end{figure}
%%%%%%%%%%%%%%%%%%%%%%%%%%%%%%%%%%%%%%%%%%%%%%%%%%%%%%%%%%%%%%%%%%%%%%%%%%

The classical
investigations that we have developed are of interest in applications to
nonlinear science, as for instance in the line of the work presented in
Refs.~{\cite{96,99,00,01}}, where one uses field theory models to mimic
nonlinear interactions in polimeric chains. For instance, the model 1 may
be used as an alternative to the $\phi^4$ model that is standardly considered
to mimic the polyacethylene (PA) chain, where Peierls instability appears due
to single and double bound alternation in carbon atoms along the
chain. The standard scenario leads to the very nice picture in which the
distance carbon-carbon is maped to the $\phi^4$ model with spontaneous
symmetry breaking. The need for spontaneous symmetry breaking is to reproduce
the two degenerate states, which describe single-double and double-single
alternations in the trans or zig-zag PA chain. Of course, this picture can be
more interesting if one adds fermions to the system, via the standard Yukawa
coupling. The alternative that we propose is to mimic the PA chain with
model 1, which is very much similar to the $\phi^4$ model at the classical
level. We hope to explore this and similar ideas in the near future.

In order to further explore the analogy between standard field theory models
and the models here introduced, we have also calculated the thermal effects
at the one-loop level. In particular, we have explored the second model,
working out the one-loop finite temperature correction to see how
the semiclassical contributions may change the classical picture of the model.
As we have shown, the finite temperature effects add to the
classical potential of model 2 and change it in a way such that
it allows for a second order phase transition to take place at high
temperature, although the symmetry is never restored by the thermal
corrections.

The potential of the second model is polynomial, and goes up to the fourth
order power in the field. In this sense it reminds us of the $\phi^4$ model.
However, it admits two degenerate but different phases, the symmetric phase
that is governed by the vanishing minimum, and the asymmetric phase which
is described by the two non-vanishing minima. In this sense it is similar
to the $\phi^6$ model. However, at the semiclassical level the thermal
corrections change the behavior of the model in a way such that it is
very specific, reminding us of neither the $\phi^4$ nor the $\phi^6$ models
anymore. 

A direct motivation that follows from the present investigations concerns
the inclusion of fermions, to see how the fermions change the scenario
we have just obtained. Another motivation concerns generalization of the
field theory models to the case of complex fields, which
give rise to models engendering the continuum $U(1)$ symmetry. In this
new scenario the models admit the presence of vortices, global and local,
depending on the gauging of the global symmetry that appears when one
changes the real field to a complex one. Furthermore, the two-field model
investigated in Sec.~{\ref{cla}} is of direct use to describe interactions
between two BEC's, and the corresponding vector solution describes the
interface between the interacting condensates, as a good alternative to
the recently proposed model of Ref.~{\cite{cha}} since in our model the
center of the interface is asymmetric, containing different quantities
of each one of the two condensates that form the interface. This is more
realist than the case studied in \cite{cha}, where the center of the interface
contains equal portions of each one of the two condensates. All the models
here introduced are also of interest to pattern formation in spatially
extended, periodically forced systems, governed by the Ginzburg-Landau
equation. This is a new route, different from the one proposed in \cite{96},
and further investigated in \cite{99,00,01}, which deals with solitons
in ferroelectric cristals, in polyethylene, and in Langmuir films. The
Ginzburg-Landau equation is appropriate for investigating fronts and
interfaces, and their contributions to pattern formation. These and other
specific issues are presently under investigation, and we hope to report
on them in the near future.

We thank M. Bahiana, P.L. Christiano, J.R.S. Nascimento, R.F. Ribeiro,
and C. Wotzasek for interesting discussions, and CAPES, CNPq, PROCAD and
PRONEX for financial support.


\begin{thebibliography}{99}
\bibitem{esc81}A.H. Eschenfelder, {\it Magnetic Bubble Technology} 
(Springer, Berlin, 1981).
\bibitem{ku84}Y. Kuramoto, {\it Chemical Oscillations, Waves, and Turbulence}
(Springer, Berlin, 1984).
\bibitem{sle98}B.A. Strukov and A.P. Levanyuk, {\it Ferroelectric
Phenomena in Crystals} (Springer, Berlin, 1998).
\bibitem{wal97}D. Walgraef, {\it Spatio-Temporal Pattern Formation}
(Springer, New York, 1997).
\bibitem{raj82}R. Rajaraman, {\it Solitons and Instantons}
(North-Holland, Amsterdam, 1982).
\bibitem{ktu90}E.W. Kolb and M.S. Turner, {\it The Early Universe}
(Addison-Wesley, Redwood, CA, 1990).
\bibitem{vsh94}A. Vilenkin and E.P.S. Shellard, {\it Cosmic Strings and 
Other Topological Defects} (Cambridge, Cambridge, UK, 1994).
\bibitem{mke}R. MacKenzie, Nucl. Phys. B {\bf303}, 149 (1998).
\bibitem{mor}J.R. Morris, Phys. Rev. D {\bf52}, 1096 (1995).
\bibitem{brs}D. Bazeia, R.F. Ribeiro, and M.M. Santos, Phys. Rev.
D {\bf54}, 1852 (1996).
\bibitem{97}F.A. Brito and D. Bazeia, Phys. Rev. D {\bf56}, 7869 (1997).
\bibitem{98}J.D. Edelstein, M.L. Trobo, F.A. Brito, and D. Bazeia,
Phys. Rev. D {\bf57}, 7561 (1998).
\bibitem{mor98}J.R. Morris, Int. J. Mod. Phys. A {\bf13}, 1115 (1998).
\bibitem{bbb99}D. Bazeia, H. Boschi-Filho, and F.A. Brito, J. High
Energy Phys. {\bf04}, 028 (1999).
\bibitem{99a}G.W. Gibbons and P.K. Townsend, Phys. Rev. Lett.
{\bf83}, 1727 (1999).
\bibitem{99b}P.M. Saffin, Phys. Rev. Lett. {\bf83}, 4249 (1999).
\bibitem{99c}H. Oda, K. Naganuma, and N. Sakai, Phys. Lett.
B {\bf471}, 148 (1999).
\bibitem{00a}D. Bazeia and F.A. Brito, Phys. Rev. Lett. {\bf84},
1094 (2000).
\bibitem{00b}D. Bazeia and F.A. Brito, Phys. Rev. D {\bf61},
105019 (2000).
\bibitem{00c}M. Shifman and T. ter Veldhuis, Phys. Rev. D
{\bf62}, 065004 (2000).
\bibitem{agm}A. Alonso Izquierdo, M.A. Gonzalez Leon, and J. Mateos Guilarte
Phys. Lett. B {\bf480}, 373 (2000).
\bibitem{h}R. Hofmann, Phys. Rev. D {\bf62}, 065012 (2000).
\bibitem{00d}D. Bazeia and F.A. Brito, Phys. Rev. D {\bf62},
101701(R) (2000).
\bibitem{bv01}D. Binosi and T. ter Veldhuis, Phys. Rev. D
{\bf63}, 085016 (2001).
\bibitem{bb01}F.A. Brito and D. Bazeia, Phys. Rev. D {\bf64},
065022 (2001).
\bibitem{n}M. Naganuma and M. Nitta, Prog. Theor. Phys. {\bf 105},
501 (2001).
\bibitem{nns01}M. Naganuma, M. Nitta, and N. Sakai, Phys. Rev. D {\bf65},
045016 (2002).
\bibitem{s1}J.H. Schwartz, Nucl. Phys. B ({\it Proc. Suppl.})
{\bf55}, 1 (1997).
\bibitem{s2}J. Maldacena, Adv. Theor. Math. Phys. {\bf2}, 231 (1998).
\bibitem{s3}A. Giveon and D. Kutasov, Rev. Mod. Phys. {\bf71},
983 (1999).
\bibitem{96}D. Bazeia, R.F. Ribeiro, and M.M. Santos,
Phys. Rev. E {\bf54}, 2943 (1996).
\bibitem{99}D. Bazeia and E. Ventura, Chem. Phys. Lett. {\bf303},
341 (1999).
\bibitem{00}E. Ventura, A.M. Simas, and D. Bazeia, Chem. Phys. Lett.
{\bf320}, 587 (2000).
\bibitem{01}D. Bazeia, V.B.P. Leite, B.H.B. Lima, and F. Moraes, Chem.
Phys. Lett. {\bf340}, 205 (2001).
\bibitem{prl}B. Denardo {\it et al.}, Phys. Rev. Lett. {\bf68}, 1730 (1992).
\bibitem{nat}M. Torres, J.P. Adrados, and F.R. Montero de Espinosa, Nature,
{\bf398}, 114 (1998).
\bibitem{cou90}P. Coullet, J. Lega, B. Houchmanzadeh, and J. Lajzerowicz,
Phys. Rev. Lett. {\bf65}, 1352 (1990).
\bibitem{ce90}P. Collet and J.-P. Eckmann, {\it Instabilities and Fronts in
Extended Systems} (Princeton, Princeton/NJ, 1990).
\bibitem{ehm98}C. Elphick, A. Hagberg, and E. Meron, Phys. Rev. Lett.
{\bf80}, 5007 (1998); Phys. Rev. E {\bf59}, 5285 (1999).
\bibitem{be98}G. Bertotti, {\it Hysteresis in Magnetism}
(Academic, San Diego/CA, 1998).
\bibitem{b}E.B. Bogomol'nyi, Sov. J. Nucl. Phys. {\bf24}, 449 (1976).
\bibitem{ps}M.K. Prasad and C.M. Sommerfield, Phys. Rev. Lett.
{\bf35}, 760 (1975).
\bibitem{bms01a}D. Bazeia, J. Menezes, and M.M. Santos,  Phys. Lett.
B {\bf521}, 418 (2001).
\bibitem{bms01b}D. Bazeia, J. Menezes, and M.M. Santos, Nucl. Phys. B
{\bf636}, 132 (2002).
\bibitem{loh79}M.A. Lohe, Phys. Rev. D {\bf20}, 3120 (1979).
\bibitem{abl01}C.A.G. Almeida, D. Bazeia, and L. Losano, J. Phys.
A {\bf34}, 3351 (2001).
\bibitem{k98}Y.S. Kivshar, D.E. Pelinovsky, T. Cretegny, and M. Peyrard,
Phys. Rev. Lett. {\bf80}, 5032 (1998).
\bibitem{95}D. Bazeia, M.J. dos Santos, and R.F. Ribeiro,
Phys. Lett. A {\bf208}, 84 (1995).
\bibitem{hkd77}B. Horovitz, J.A. Krumhansi, and E. Domany, Phys.
Rev. Lett. {\bf14}, 778 (1977).
\bibitem{tdl79}S.E. Trullinger and R.M. DeLeonardis, Phys. Rev.
A {\bf20}, 2225 (1979).
\bibitem{the99}S. Theodorakis, Phys. Rev. D {\bf60}, 125004 (1999).
\bibitem{shi98}M.A. Shifman and M.B. Voloshin, Phys. Rev. D {\bf57},
2590 (1998).
\bibitem{mya}C.J. Myatt {\it el al.}, Phys. Rev. Lett. {\bf78},
586 (1997).
\bibitem{tim}E. Timmermans, Phys. Rev. Lett. {\bf81}, 5718 (1998).
\bibitem{cha}S. Coen and M. Haelterman, Phys. Rev. Lett.
{\bf87}, 140401 (2001).
\bibitem{baz99}D. Bazeia, Phys. Rev. D {\bf60}, 067705 (1999).
\bibitem{j74}R. Jackiw, Phys. Rev. D {\bf9}, 1686 (1974).
\bibitem{dj74}L. Dolan and R. Jackiw, Phys. Rev. D {\bf9}, 3320 (1974).
\bibitem{b83}D. Bazeia, G.C. Marques, and I. Ventura, Rev. Bras. Fis.
{\bf 13}, 253 (1983).
\bibitem{a85}C. Arag\~ao de Carvalho {\it et al.}, Phys. Rev. D
{\bf31}, 1411 (1985).
\bibitem{mse79}R.N. Mohapatra and G. Senjanovi\v c, Phys. Rev.
D {\bf20}, 3390 (1979).
\end{thebibliography}
\end{document}